# Edge pinning and transformation of defect lines induced by faceted colloidal rings in nematic liquid crystals

Bohdan Senyuk,[1] Qingkun Liu,[1] Ye Yuan,[1] and Ivan I. Smalyukh[1,2,3,4,*]

[1]*Department of Physics, University of Colorado, Boulder, Colorado 80309, USA*

[2]*Department of Electrical, Computer, and Energy Engineering, University of Colorado, Boulder, Colorado 80309, USA*

[3]*Soft Materials Research Center and Materials Science and Engineering Program, University of Colorado, Boulder, Colorado 80309, USA*

[4]*Renewable and Sustainable Energy Institute, National Renewable Energy Laboratory and University of Colorado, Boulder, Colorado 80309, USA*

*Email: ivan.smalyukh@colorado.edu

**Abstract**

Nematic colloids exhibit a large diversity of topological defects and structures induced by colloidal particles in the orientationally ordered liquid crystal host fluids. These defects and field configurations define elastic interactions and medium-mediated self-assembly, as well as serve as model systems in exploiting the richness of interactions between topologies and geometries of colloidal surfaces, nematic fields, and topological singularities induced by particles in the nematic bulk and at nematic-colloidal interfaces. Here we demonstrate formation of quarter-strength surface-pinned disclinations, as well as a large variety of director field configurations with splitting and reconnections of singular defect lines, prompted by colloidal particles with sharp edges and size large enough to define strong boundary conditions. Using examples of faceted ring-shaped particles of genus $g = 1$, we explore transformation of defect lines as they migrate between locations in the bulk of the nematic host to edge-pinned locations at the surfaces of particles and vice versa, showing that this behavior is compliant with topological constraints defined by mathematical theorems. We discuss how transformation of bulk and surface defect lines induced by faceted colloids can enrich the diversity of elasticity-mediated colloidal interactions and how these findings may impinge on prospects of their controlled reconfigurable self-assembly in nematic hosts.

# I. INTRODUCTION

Liquid crystals (LCs) and colloids are two principal subclasses of soft condensed matter with both fundamental science and technological significance [1], but without a well-defined boundary separating them as distinct material systems. Indeed, colloidal dispersions of anisotropic particles often form LC phases [1], LCs often serve as fluid hosts of colloidal dispersions [2–9], and so on. For the LC colloids comprised of particles dispersed in a nematic fluid, a property of particular interest is the particle-induced defects and elastic distortions that accompany solid or fluid colloidal inclusions, predetermining medium-mediated interparticle forces, self-assembly, and response to external stimuli such as electric fields [4,10–19]. A large variety of defect configurations can be induced by spherical colloidal particles in the nematic LC host, including bulk point defects such as hyperbolic hedgehogs [3,4,9,10], surface point defects dubbed "boojums" [10], and half-integer disclination rings often called "Saturn ring" defects [7–9], in which such defects emerge to help match surface boundary conditions on particle surfaces with an incompatible far-field LC director. Nonspherical colloidal particle shapes allow for enriching the diversity of topological defect generation in the LC host even further [19–31]. For example, by exploiting relations between topologies of surfaces and fields prescribed by topological theorems [25,28,30] one can define the net hedgehog topological charge of induced point and ring-shaped defects in the bulk of LC and at LC-particle interfaces. Another example is the ability to define the location of defect rings shifted away from the polygonal platelet's midplane by controlling the geometry of truncated pyramids [31]. However, these and other examples of using colloidal particle geometric shapes to control the particle-induced defects present only a few approaches out of a wide variety of possibilities that remain to be exploited.

In this work, we uncover a regime of interaction between the surface boundary conditions on geometrically and topologically nontrivial colloidal particles and the surrounding nematic LC host, which yields a series of defect and director structures that have not been observed in LC-colloidal systems previously. The key feature of the studied colloidal particles is that the smallest particle's dimension is much larger than the surface anchoring extrapolation length in the range of 100–500 nm, which yields strong surface boundary conditions that cannot be violated even at sharp edges of faceted particles. In addition to commonly observed half-integer defect lines encircling the particles, we find quarter-strength disclinations pinned to edges of the faceted particles, as well as nodes of defect lines with different strengths, of which some are pinned to colloidal surfaces while others are bulk defect lines. Using conventional and nonlinear optical microscopy as well as holographic optical tweezers [32–34], we probe director field and defect configurations, demonstrating interrelations between the particle topology and the properties of the induced defects. Our experimental findings demonstrate that the strength (winding number) of individual surface-pinned disclinations around colloidal particles is unconstrained and can be controlled by the geometry of these colloidal inclusions, while the total topological characteristics, such as the strength of defect lines and the topological hedgehog charge of defect rings remain compliant with topological theorems. Finally, we discuss how our findings may broaden the richness of phenomena in LC colloids, as well as enable alternative forms of mesostructured self-assembly.

## II. MATERIALS AND TECHNIQUES

Toroidal or ring-shaped colloidal particles were fabricated from silica ($SiO_2$) to have a larger size than what we used in our previous studies [28,30]. First, a 5-μm-thick $SiO_2$ layer was deposited on a silicon wafer using plasma-enhanced chemical vapor deposition. Then a chromium layer (∼200 nm) was sputtered onto this $SiO_2$ layer as a hard etching mask, followed by spin-coating on the top a 1.5-μm-thick layer of photoresist, AZ5214 (Clariant AG). The pattern of large toroidal particles was defined by a direct laser writing system, DWL 66FS (Heidelberg Instruments, Germany) with a semiconductor laser at 405 nm and then transferred from the photoresist layer to the chromium hard mask via wet etching process, followed by inductively coupled plasma (ICP) etching to obtain silica tori on the silicon layer. Finally, the toroidal colloidal particles were released from the supportive silicon wafer by selectively ICP etching the silicon. The fabricated solid tori-shaped particles had an outer diameter of approximately 25 μm [Figs. 1(a)–1(d)] with a square-shaped (∼5 × 5 $μm^2$) cross section of the torus tube and four circular edges at which the side and top or bottom faces of the colloidal ring come together at 90°. Strong homeotropic boundary conditions for a director field **n**(**r**) of a nematic LC 4-cyano-4′-pentylbiphenyl (5CB, from Frinton Laboratories, Inc.) were imposed by treating the colloidal tori with an aqueous solution (0.05 wt %) of N,N-dimethyl-N-octadecyl-3-aminopropyl-trimethoxysilyl chloride (DMOAP). Dispersions of the tori in the nematic LC were infiltrated in-between two glass plates separated by calibrated glass spacers (Duke Scientific Corp.) defining a gap thickness of used samples, which was varied in the range from ∼15 to 60 μm. Before cell assembly, substrates were treated with DMOAP to achieve a perpendicular far-field director $n_0$ or spin coated with a polyimide PI2555 (HD Microsystems) for homogeneous in-plane alignment of $\mathbf{n}_0$ enforced by unidirectional rubbing of films after baking polyimide at 270 °C for 1 h. One of the substrates in the cell design was 150-170 μm in thickness to minimize spherical aberrations and to comply with short working distances during the imaging and optical manipulation experiments involving high numerical aperture (NA) immersion oil objectives.

We used a multimodal experimental setup built around an inverted Olympus IX81 microscope for conventional bright-field and polarizing optical microscopy observations, multiphoton excitation fluorescence polarizing microscopy, and holographic optical manipulations. A tunable (680–1080 nm) Ti:sapphire oscillator (140 fs, 80 MHz, Chameleon Ultra-II, Coherent) was used for the multiphoton excitation fluorescence polarizing imaging [27,33]. Fluorescence images of $\mathbf{n}(\mathbf{r})$ in 5CB were obtained in a three-photon excitation fluorescence polarizing microscopy (3PEF-PM) mode utilizing a three-photon excitation of 5CB molecules by femtosecond laser light at 870 nm; the resulting fluorescence signal was detected within a spectral range of 387–447 nm by a photomultiplier tube H5784-20 (Hamamatsu) [27]. The in-plane position of a focused excitation beam was controlled by a galvano-mirror laser-scanning unit (FV300, Olympus). Polarization of the excitation light was varied using a half-wave plate mounted immediately before a 100× (NA = 1.42) oil immersion objective. Optical microscopy textures of colloidal tori were recorded with a CCD camera (Flea, Point Grey Research, Inc.) and were used in conjunction with 3PEF-PM imaging to determine $\mathbf{n}(\mathbf{r})$ and defect configurations in our experimental soft matter system. Optical manipulations of toroidal particles and defects around them were realized with a holographic optical trapping (HOT) system [27,34] operating at a wavelength of $\lambda = 1064$ nm and using the same objectives as for the conventional optical and nonlinear 3PEF-PM imaging.

## III. RESULTS

Despite their relatively large size, specially designed particles used in our experiments are still Brownian colloidal particles stabilized against sedimentation in the bulk of a LC host [Figs. 1(a)–1(d)] by the balance between forces of gravity and strong elastic repulsions from confining substrates of a LC cell. Toroidal particles with homeotropic surface anchoring dispersed in a nematic LC deform the initially uniform director field $\mathbf{n}(\mathbf{r}) = \mathbf{n}_0$ and typically cause formation of singular point or line topological defects [28], albeit only line defects will appear in the field configurations studied in the present work. The strength $k = \pm\theta/(2\pi)$ of line defects or disclinations is determined by an angle $\theta$ by which the director rotates along a path encircling the defect's singular core, with a sign defined by a sense of **n(r)** rotation relative to the navigation direction [35]. The nonpolar nature of the nematic LC constrains the smallest value of the strength of the bulk defect lines to be $|k| = 1/2$, as well as allowed values to be $|k| = Q/2$, where $Q$ is an integer number, although these constraints do not apply to surface-bound defect lines, as we will discuss below. The high-strength singular defect lines in the LC bulk with $|k| > 1/2$ tend to split to the elementary-strength half-integer defect lines of the same net strength but of a lower elastic free energy. The half-integer defect lines $k = \pm 1/2$ are the only topologically stable defect lines in three-dimensional (3D) nematic LCs, topologically nondistinguishable from each other because of being able to continuously transform from a $k = 1/2$ line to the one with $k = -1/2$ and vice versa. We will use the strength k of defect lines as defined above to characterize the local structure of both bulk and surface-bound defect lines in the cross sections orthogonal to them.

In the most conventional (or the ground state) field configuration induced by the colloidal tori, which was previously extensively explored for smaller ring-shaped particles with cross-section dimensions comparable to the surface anchoring extrapolation length, two disclinations of $k = -1/2$ span along the inner and outer sides of a torus, forming closed loops often called "Saturn ring" defects [Figs. 1(a), 1(c), 1(f), and 1(h)]. In this configuration, a normal **s** to a plane containing the torus is parallel to $\mathbf{n}_0$ and the **n(r)** structure is symmetric with respect to the particle's midplane [Figs. 1(a), 1(c), 1(f), and 1(h)] in both homeotropic and planar cells. For **s**$\|\mathbf{n}_0$, the minimization of distortions of mirror-symmetric deformations of **n(r)** above and below the bulk defect rings stabilizes them at a midheight of a torus tube far from the edges [Figs. 1(a), 1(f), and 1(h)]. Polarizing optical microscopy textures and 3PEF-PM imaging demonstrate that distortions of **n(r)** in this configuration propagate from the particle by large distances comparable to the width of the tube of the toroidal particle [see Fig. 1(c) and the inset in Fig. 1(a)]. The discontinuities at the particle's edges can be neglected in the analysis of defect structure and transformation when a torus tube has rounded edges or its entire cross section's dimension is comparable or smaller than the surface anchoring extrapolation length [28]. However, sharp geometrical discontinuity and strong incompatibility of boundary conditions at the edges of relatively large particles, such as the ones used in the present study, lower the scalar order parameter at the edges and lead to formation of singular surface-bound defect lines [23,26,36,37]. Close inspection of 3PEF-PM cross-sectional images [Figs. 1(c) and 1(d)] allows one to conclude that the edges of the studied toroidal particles are sharp. The nanometer-scale roundness of these sharp edges certainly exists as the edges are not atomically sharp, but the radius of curvature of the edges is orders of magnitude smaller than the surface anchoring extrapolation length of the LC, which makes it insignificant in the present experiments.

We show below that details of **n**(**r**) deformations and the ensuing surface-bound defects at the sharp edges of large faceted colloidal particles should be taken into account in the topological analysis of director structures and can cause unexpected and complex transformation of bulk and surface defect lines as well as the free-energy-minimizing alignment of colloidal rings at **s**∦**n**$_0$ [Figs. 1(b), 1(d), 1(e), 1(g), and 1(i)]. Figure 1(b) shows an example of colloidal-ring-induced field configuration for a colloidal ring in a homeotropic cell with **s** tilted with respect to **n**$_0$. The tilted colloidal rings are frequently observed configurations in the nematic LC cells. This tilt can be spontaneous due to minimization of the elastic free energy (with the minima corresponding to metastable or stable states) or induced by the confinement. Spontaneously tilted tori are observed in both homeotropic and planar cells with a thickness $d$ larger than the corresponding dimensions of the particles: $d$ has to be larger than the width of the tube in the homeotropic cells [Fig. 1(d)] and larger than the diameter of the particle in the planar cells. In the planar cells, the tilt of the toroidal particles can also originate from the tight flat-cell confinement when $d$ is smaller than the diameter of the ring-shaped particle. Elastic repulsion of opposite sides of particles with homeotropic surface anchoring from the substrates with incompatible planar surface alignment forces particles to tilt with respect to the far-field director.

In the case of tilted particles, rather than observing bulk ring-shaped disclination loops, we find four handle-shaped branches of bulk defect lines bound to the particle surfaces at the sides of the torus. In the optical micrographs, these bulk defect lines are seen as kinklike protrusions at both the inner and outer faces of the particles [Fig. 1(b)]. The locations of these handle-shaped defect lines and the sites at which they are anchored to the particle surface can be more or less aligned along a straight line corresponding to the particle's tilt axis [Figs. 1(b) and 1(d)] or, alternatively, they can be displaced with respect to each other and this axis, as shown in Fig. 1(e). The details of the 3D director field deformations (Figs. 1–4) can be analyzed using optical microscopy and 3PEF-PM imaging. One can establish that a kinky protrusion visible at the side of the torus [Figs. 1(b), 1(e), 2(a), 2(b), and 2(d)] is a defect line traversing from the top to the bottom edge of the colloidal ring through the LC bulk [Fig. 2(e)]. These handlelike disclination fragments can be seen spanning between the edges of both the inner and outer faces of the colloidal ring. This configuration is very different from that of mutually parallel ring-shaped defect lines winding around the outer and inner faces that are seen in Fig. 1(a). Another interesting observation as compared to the nontilted particles is the distance to which the optically detectable $\mathbf{n}(\mathbf{r})$ deformations propagate away from the particle surfaces. This distance is significantly shorter for the tilted particles, so that the director distortions are mostly localized at the edges of the particles [compare Figs. 1(c) and 1(d) and the insets in Figs. 1(a) and 1(b)] and the kinklike protrusions extend beyond them. Half-integer defect lines cannot terminate in the bulk of LC because they are topologically distinct from its uniform ground-state structure, a constraint that generally does not hold at the LC-colloidal surfaces. Moreover, the fact that the fragments of defect lines terminate at the edges of particles with well-defined boundary conditions points towards transformation of bulk and surface defects, which we explore below.

The sharp change of the orientation of the faces of the square cross section of a colloidal ring with respect to $\mathbf{n}_0$ introduces incompatible boundary conditions, thus diminishing the nematic degree of order at the particle edges [9,23,36,37,38]. This incompatibility of director orientations in the LC bulk can be resolved via deformations where the director $\mathbf{n}(\mathbf{r})$ rotates by $\theta = \pm\pi/2$ around the edge, resulting in a surface defect line of a strength $k = \pm 1/4$ that we will call a "surface quarter disclination" [Figs. 1(g)–1(n)] from now on. When $\mathbf{s}\|\mathbf{n}_0$, there are two faces, top and bottom, oriented so that the imposed homeotropic boundary conditions are compatible with $\mathbf{n}_0$ and two side faces, inner and outer, with their boundary conditions orthogonal to $\mathbf{n}_0$ [Figs. 1(f), 1(h), and 1(j)–1(l)], which can result in two configurations of $\mathbf{n}(\mathbf{r})$ around the square tube of the torus, one with surface defects only [Figs. 1(j) and 1(k)] and another with surface and bulk defects at the same time [Figs. 1(f)–1(i), 1(l), and 1(m)] [23,36,37]. The first configuration, never observed in our experiments (likely due to its higher free energy cost), does not require the presence of bulk defects at all and only four surface quarter-strength disclinations, two pairs of opposite sign, are necessary to fit the particle's boundary conditions with the uniform $\mathbf{n}_0$ with the help of minimum-cost elastic distortions [Figs. 1(j) and 1(k)]. The quarter disclinations of the same sign can reside at the edges of the same flat face, top or bottom [Fig. 1(j)], or they can be positioned on diagonally opposite edges of the particle's square cross section [Fig. 1(k)]. Another, the most frequently observed configuration [Fig. 1(a)], consists of two bulk $k = -1/2$ disclination loops and four $k = +1/4$ edge disclinations [Figs. 1(f), 1(h), and 1(l)], which vanish for particles with rounded edges or relatively weak surface boundary conditions, or when the dimensions of the colloidal ring's cross section become comparable to the surface anchoring extrapolation length. The total topological hedgehog charge of the particle-induced defect structures is equal to zero in both **n**(**r**) configurations, consistent with the predictions of topological theorems for the defects induced by genus-one colloidal surfaces [28]. The conservation of topological features characterizing the observed defects can also be analyzed in cross-sectional planes containing $\mathbf{n}_0$ and crossing the ring-shaped particle along with the defects that it induces, such as the cross-sectional planes shown in Figs. 1(f)–1(n). For all these structures, the strengths due to defects and the particle's boundary conditions add to zero, consistent with the uniform alignment of the director away from the particle [Figs. 1(f)–1(n)].

The structure of observed defects changes dramatically with tilting of the faceted colloidal tori, when the alignment of particles is such that **s**∦$\mathbf{n}_0$ [Figs. 1(b), 1(d), and 1(e)]. The square cross section of the torus's square-shaped tube rotates [Figs. 1(m) and 1(n)] upon tilting the torus and enforced alignment at its faces changes their orientation with respect to $\mathbf{n}_0$. The bulk negative half-integer disclinations are still stabilized in between the top and bottom edges of the particle but are sliding along the torus side and approaching one of the side edges with a positive quarter-strength defect line at the edge [Fig. 1(m)]. Eventually, at some tilt angle $\alpha \approx 5°$ − 20° between **s** and $\mathbf{n}_0$ [Fig. 1(d)], the inner and outer $k = -1/2$ bulk defect lines get absorbed into the edges changing their effective strength to $k = -1/4 = -1/2 + 1/4$ through the defect transformations at the nodes [Fig. 1(n)]. This is consistent with the signs of quarter-strength line defects at the edges of particles in homeotropic cells shown in Figs. 1(g) and 1(i) using different coloring. The 3PEF-PM imaging [Fig. 1(d)] indicates that small distortions are localized at the edges of the particle. As the scanning with an excitation laser beam was performed from the direction of the bottom substrates one can see the signal corresponding to the distortions only at the bottom edges because the signal from the top edges and above the particle is scattered and partially blocked by the particle as evidenced by a dark area above the particle. The symmetry of the detected small distortions is different at the right and left sides of the torus tube cross sections, which can be attributed to the opposite sign of the defects. The confocal and multiphoton-process-based 3D imaging techniques are known to be sensitive to nonuniformities in refractive index within the studied medium such as our LC-colloidal composite, yet they are powerful tools to study them. The important feature to notice here is that, upon tilting the torus, the far ends of both outer and inner bulk disclinations with respect to a tilt axis absorb to opposite – top and bottom – edges of the colloidal torus [Figs. 1(g), 1(l), 1(m), and 3]. These newly formed structures with negative quarter defects, both at the inner and outer sides, propagate along opposite circular edges towards the middle of the torus to a branching point/node where they desorb from the edge and continue as the positive quarter defect and the bulk half-integer defect line jumping from the top to bottom edges (Fig. 3), as can be deduced from the experimental observations [Figs. 1(b) and 2(e)]. Interestingly, each of the four circular edges has two quarter disclination arcs of an opposite sign meeting at the branching points connected by bulk half-integer disclination fragments (Fig. 3). The typical length of these bulk fragments is about the side width $a \approx 5$ μm of a square-shaped tube forming the ring [Fig. 2(e)]. Topological transformation at the branching node conserves the strength of defect lines: the $k = -1/4$ surface disclination on one side of the node transforms into a surface line with $k = +1/4$ and a bulk disclination with $k = -1/2$ after the node, so that $-1/4 = -1/2 + 1/4$.

Similar reversible transformation of bulk half-integer disclinations into surface-bound quarter disclinations at particle edges of a tilted faceted colloidal ring (and vice versa) happens also in planar and twisted nematic cells. Figure 4 shows a tilted torus in the bulk of a planar LC cell. The propagation and transformation of defect lines is similar to the ones in a homeotropic cell (Fig. 1), but orientation of the defect segments with respect to confining substrates changes by $\sim\pi/2$ and quarter defects at the edges pointing towards confining substrates have negative strengths in this case (Fig. 4). The elastic deformations and defect structures in a twisted nematic cell are even more complex. We used a twisted nematic cell of a thickness $d$ comparable or larger than a torus diameter. In addition to the out-of-plane tilting the torus experiences the in-plane rotation due to the weak twist of LC (the pitch of the confinement- and boundary-conditions-induced helicoidal structure is $p = 4d \gtrsim 100$ μm). Figure 2 shows a tilted faceted torus in a twisted nematic cell of $d \approx (25\text{-}30)$ μm. The height $h$ of the center of the torus location with respect to the bottom substrate is determined by the balance of the forces of gravity and elastic repulsion from both substrates. Therefore the torus is rotated accordingly, so that the in-plane projection of **s** is parallel to a local director at $h$ that is $s_\perp \| \mathbf{n}(h)$. An interesting feature is that the sense of rotation of the colloidal ring (and $s_\perp$) with respect to an easy axis $\mathbf{e}_b$ at the bottom substrate is the same as the sense of twisting of the LC [Fig. 2(b)]. This peculiarity can be used to determine the handedness of twist in the LC confined into a twisted cell that contains $\pm\pi/2$ twisted domains: If $s_\perp$ of the tilted torus rotates counterclockwise with respect to $\mathbf{e}_b$ then the twist is right-handed; in contrast, the twist within the region is left-handed when rotation is clockwise. Bulk disclination fragments on the inner and outer faces of the torus in the twisted cell are arranged along the tilt axis perpendicular to $s_\perp$ [Figs. 2(a)–2(c)]. What is interesting is that due to the twist and thus different orientation of a local $\mathbf{n}(\mathbf{r})$ at the top and bottom (with respect to substrates) parts of the torus, the bulk disclination can also be aligned along a straight line parallel to $s_\perp$ and perpendicular to the tilt axis [Figs. 2(d)–2(f)]. Owing to this orientation of bulk defect lines, one can clearly see on the colloidal flat face pointing in the observation direction that the bulk disclination is jumping across the torus's face from one of its edges to the other [Figs. 2(d) and 2(e)].

To investigate the strength of absorption of bulk half-integer defect lines into the edges of faceted colloidal particles, we used optical manipulations with holographic optical tweezers (Fig. 5). In the first experiment, a high-power optical beam was used to melt the LC locally around the tilted toroidal particle in a nematic cell. To facilitate the local melting of the LC to an isotropic phase (*I*) with a high-power trapping laser beam, we used confining cell substrates with indium-tin-oxide (ITO) films, which absorb laser light at 1064 nm and convert it to heat. After the trapping beam and the ensuing heating process are turned off, a melted LC around a particle [Fig. 5(a)] is quickly quenched back to the nematic phase (*N*) and one can observe transient processes of nucleation and transformation of thin dark lines corresponding to bulk half-integer disclinations [35]. Long entangled disclination lines appear immediately after the local *I*–*N* transition, and transform and wind around the torus while getting absorbed into the edges of the particle [Figs. 5(b)–5(f)]. Figures 5(b)–5(f) show a time sequence of bulk disclinations being absorbed into the particle's edges. This process completes within hundreds of milliseconds, which is a typical characteristic time of spontaneous relaxation of nematic **n**(**r**) distortions in LC samples of the size used in our experiments [35]. As a result of this process, only short segments of bulk disclinations are left at the particle's sides, inside and outside the torus, traversing between the top and bottom edges [Fig. 2(e)]. Occasionally, additional dark or bright spots randomly distributed on the surface of the particles can be found in the optical microscopy textures. They correspond to the residual debris present in the samples due to the local imperfections of the particles or different solid residuals that remain after transferring particles from the silica substrates to the LC. However, the size of the debris is much smaller than the dimensions of the colloidal particles and their presence does not affect the director and defect configurations. We have also used small isotropic droplets, produced by local melting of 5CB with a high-power (150-200 mW) laser beam due to the absorption of light, to trap and stretch these fragments of bulk disclinations [18,39] on the outer and inner faces of the torus [Figs. 5(g)–5(i)]. Small isotropic droplets allow strong trapping of $k = -1/2$ bulk disclinations and their manipulation with optical tweezers. Peeling the defect lines off the edges would require desorbing them from the edges, which requires locally transforming a negative quarter-strength surface defect line into a positive quarter-strength surface defect and a negative half-integer bulk disclination. However, trapped disclinations inside and outside the torus remain pinned at the branching points during stretching by the laser tweezers. The exerted pulling force of the laser trap, which extends the handles of the defect lines further into the LC bulk, and the ensuing defect line tension of the half-integer disclination are insufficient to peel them off the edges and to mediate the necessary topological transformations of $\mathbf{n}(\mathbf{r})$. This is consistent with our original observations described above, which reveal that bulk defect lines strongly absorb into the particle edges and transform into the $k = +1/4$ surface defect line on one side of the node at the edge and into the $k = -1/4$ surface defect line on the other side (Figs. 1 and 3).

Even though our study in this work is mainly focused on the defect structures with transformations between the bulk half-integer and surface quarter-strength disclinations, we have also observed several other defect structures, including but not limited to the ones we previously reported in Ref. [28]. For example, within one of the particle-defect configurations, the outer stretched half-integer disclination loop is following the edges of a tilted toroidal particle and transitions between the top and bottom ring-shaped edges uninterrupted (Fig. 6). This type of defect configuration somewhat resembles what was found in several recent theoretical studies of genus-zero nematic colloids with sharp edges, such as the cylinders with the flat end faces [40–42], albeit these studies dealt with particle dimensions comparable to the surface anchoring extrapolation length while the particles we study here are much larger. By examining optical micrographs, one can clearly see that $\mathbf{n}(\mathbf{r})$ distortions originating from the pinned half-integer defect loops in Fig. 6 propagate rather far into the LC bulk [compare insets in Figs. 1(a), 6(a), and 1(b)], similar to the case of more conventional [28] undistorted defect loops around a nontilted torus, in which these defect loops are parallel to the plane of the particle ring. Unlike in the case of genus-zero particles with sharp edges, the topology of our genus-one rings does not require inducing defects within the LC, but we show that many types of defect structures can nonetheless occur to match the strong boundary conditions on faceted rings to that of the uniform far-field director. The large variety of observed colloidal-defect structures additionally demonstrates that colloidal handle-body-shaped particles [28,30] suspended in an anisotropic LC host fluid can access a wide variety of metastable states and cause unusual transformations of defects not accessible to defects induced by spherical inclusions, or even geometrically complex colloidal inclusions with the simple genus-zero topology. Further, our study in this work demonstrates that the size of ring-shaped particles matters and that the particles with large cross-section dimensions relative to the surface anchoring extrapolation length can exhibit behavior different from that of small colloids with the same topology and geometric features [28].

## IV. DISCUSSION

All previous studies of LC-colloidal systems dealt with either surface or bulk particle-induced defects, but our experiments described above demonstrate that the bulk and surface defects can coexist in the LC surrounding of the same particle and can even transform from one type to another, forming nodes of bulk and surface defect lines. These observations highly enrich the interplay between the topologies of colloidal surfaces and nematic field configurations that they induce in the uniform LC background. Although we studied ring-shaped particles with edges connecting flat and curved faces at 90° with respect to each other, naturally defining the ubiquitous appearance of quarter-strength surface-bound nematic defect lines, the control of geometric shape (e.g., the control of angles of misalignment between faces at sharp edges of concave and convex particles of different geometry [26,31,43]) may allow for variable strength $k$ of the particle-induced surface defect lines. In future works, therefore, it will be even more interesting to explore how the surface-bound defect lines with unconstrained $k$, such as, for example, $k = +1/6$ or $k = -2/9$ surface defect lines, can then transform into the bulk defect lines constrained to have $|k| = Q/2$ and vice versa. It will also be of interest to explore how the strength of surface anchoring, in addition to faceted-shape geometry, can determine $k$ values and the behavior of these defects. For example, the finite surface anchoring strength can cause departures of $k$ values of disclinations pinned to the 90° edges between flat particle faces away from $k = \pm 1/4$. In general, the richness of possibilities that can arise can be modeled by placing normal nematic disclinations with $k = \pm Q/2$ and considering the k values as defined within the LC bulk that remains after the virtual part of this defect line is depressed inside the colloidal particle with a given type and strength of surface boundary conditions. This shows that even when topological characteristics of colloidal surfaces remain unchanged, much of the desired control over the particle-induced defects can be achieved by varying the geometric shape features of these LC-colloidal objects. The experimental exploration of such interplay of colloidal surface geometry, topology, and the corresponding field and defect configurations may provide insights into fields well beyond LC colloids and soft condensed matter. We foresee that similar effects can be observed for faceted colloids with sharp edges and with genus $g = 0$ and also for large-genus particles such as handle bodies with $g \gg 1$, and it will be of interest to generalize our findings to these other topologically different colloidal surfaces in future works.

The surface anchoring boundary conditions at the surface of particles incompatible with the orientation $\mathbf{n}_0$ causes distortions of a director field with deformations of $\mathbf{n}(\mathbf{r})$ propagating to distances far from the particle. Distribution of $\mathbf{n}(\mathbf{r})$ at the surface of the sphere enclosing such particles is responsible for formation of elastic multipoles, which determine weak perturbations of $\mathbf{n}(\mathbf{r})$ far from the particles and their LC-mediated interactions [2–6,8,11–16,19,20,31,43]. In our experiments, symmetry of director distortions far away from the ring-shaped particles depends on the defect configurations that we explore. For example, the structures of defects shown in Figs. 1(a), 1(f), 1(h), and 1(l) induce elastic quadrupoles in $\mathbf{n}(\mathbf{r})$, but other structures can be characterized by elastic dipoles of different types or can also have mixed elastic multipoles. We can analyze these elastic multipoles using the approach of nematostatics [14–16]. Due to the symmetry of a uniaxial nematic LC host itself, the symmetry operations over $\mathbf{n}(\mathbf{r})$ induced by our particles are restricted to rotations around $\mathbf{n}_0$ and $\pi$ rotations about any horizontal axis perpendicular to $\mathbf{s} \| \mathbf{n}_0$, and mirror reflections in any vertical plane passing through $\mathbf{n}_0$ and a horizontal plane perpendicular to $\mathbf{n}_0$. Symmetry transformations about the tilted $\mathbf{s}$ ($\mathbf{s} \nparallel \mathbf{n}_0$) are not allowed. Structures around a normal ($\mathbf{s} \| \mathbf{n}_0$) faceted torus can form an isotropic uniaxial elastic dipole with $C_{\infty v}$ symmetry when there are only surface defects $|k| = 1/4$ at the edges around a particle [Figs. 1(j) and 1(k)] or an elastic quadrupole with symmetry of a nematic $D_{\infty h}$ when there are two bulk disclinations $k = -1/2$ and four $k = +1/4$ defects encircling the particle [Figs. 1(a), 1(c), 1(f), 1(h), and 1(l)]. However, different types of elastic dipoles form around the spontaneously tilted faceted tori. In a homogeneous, infinitely thick nematic, the total elastic energy due to $\mathbf{n}(\mathbf{r})$ distortions around a tilted torus far from the confining substrates does not change upon its rotations around $\mathbf{n}_0$ or mirror reflection in a plane parallel to $\mathbf{n}_0$ and perpendicular to a tilt axis [Figs. 1(b), 1(d), 1(g), and 1(i)], which allows us to identify the ensuing elastic multipole as a biaxial nonchiral dipole with $C_{1v}$ symmetry. This elastic dipole transforms into a chiral biaxial dipole with $C_1$ in a twisted nematic cell. Tight confinement of a nematic LC can additionally lower symmetry of $\mathbf{n}(\mathbf{r})$ distortions in planar (Fig. 4) and twisted (Fig. 2) cells to $C_i$ as rotations around $\mathbf{n}_0$ become prohibited due to changing $\mathbf{n}(\mathbf{r})$ with respect to confining substrates. This simple analysis shows that a diverse variety of elastic multipoles can be achieved by varying the induced surface and bulk defects even for the particles of exactly the same geometric shape and topology. Provided that the type of particle-induced defects can be preselected, this can enable unprecedented richness of colloidal self-assembled structures mediated by the LC elasticity. On the other hand, in addition to purely elastic LC-mediated interactions, one can envisage alternative forms of self-assembly that involve the interlinking of the bulk fragments of handle-shaped defect lines of different particles as well as defect lines terminating on different particles. A large diversity of field and defect configurations is also expected for tangentially or conically degenerate [44], hybrid [38], and optically controlled [45] boundary conditions on colloidal surfaces. In addition to direct experimental explorations of such diverse forms of self-assembly, additional insights can be provided by means of numerical modeling which were already successfully applied to probe LC ordering at the edges between faceted confining and colloidal surfaces [23,36,37].

## V. CONCLUSIONS

To conclude, we experimentally demonstrate a regime of elastic interactions between the surface boundary conditions on geometrically and topologically nontrivial colloidal particles and a surrounding nematic LC host, which yields a series of defect and director structures that have not been observed in the previous experiments on the LC colloidal systems. In addition to commonly observed half-integer defect lines encircling and entangling spherical and topologically nontrivial particles [13,18], we find surface quarter-strength defect lines pinned to sharp edges of faceted particles, as well as nodes of defect lines with different strengths, of which some are pinned to colloidal surfaces while others are the bulk defect lines. Using conventional and nonlinear optical microscopy as well as holographic optical tweezers [32–34], we have probed director field and defect configurations, demonstrating interrelations between the particle topology and properties of induced defects. Our experimental findings demonstrate that the strength (winding number) of individual surface-pinned disclinations around colloidal particles is unconstrained and can be controlled by the geometry of these colloidal inclusions while the overall topological characteristics of particle-induced defects comply with topological theorems and conservation laws.

## ACKNOWLEDGMENTS

We acknowledge support of the National Science Foundation Grant No. DMR-1410735 (B.S., Q.L., Y.Y., and I.I.S.) and the Humboldt foundation (I.I.S.), as well as discussions with M. Tasinkevych. I.I.S. also acknowledges the hospitality of the Max Planck Institute for Intelligence

**Figures**

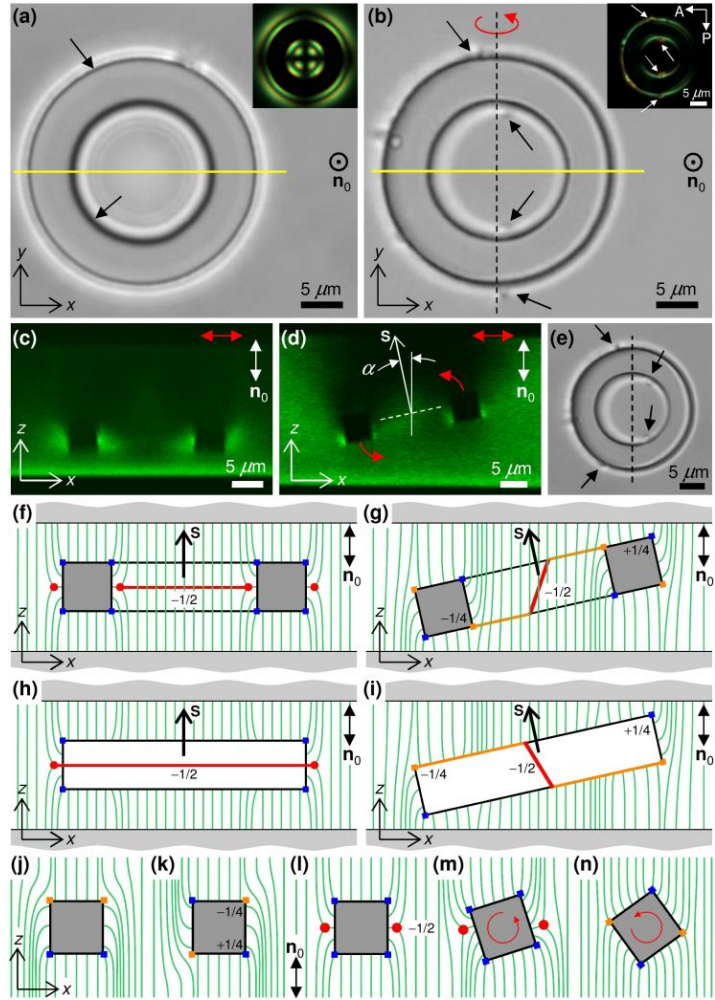

FIG. 1. Defects transformation around torus-shaped colloidal particles in a nematic LC: (a,b,e) Bright-field optical and (c,d) 3PEF-PM micrographs of an orthogonally aligned (a,c) and tilted (b,d,e) particle. Cross-sectional $xz$ images in (c,d) were obtained along a yellow line in (a,b). Insets in (a,b) show corresponding polarizing microscopy textures obtained between crossed polarizers $P$ and $A$. Double arrows and circles with a dot in the center show the in-plane and out-of-plane directions of the far-field director $\mathbf{n}_0$, respectively. Red double arrows show the direction of polarization of the excitation beam parallel to the plane of the image. Black dashed lines in (b,e) show the particle's tilt axis. (f–i) Corresponding schematic diagrams of $\mathbf{n}(\mathbf{r})$ and defects inside (f,g) and outside (h,i) the ring-shaped particle; only −1/4 defect propagating along edges is shown for clarity. A black thick arrow shows a normal $\mathbf{s}$ to the plane of the ring-shaped particle. Curved red arrows in (b,d,m,n) show the direction of tilt. White dashed line in (d) shows a plane of the ring-shaped particle and $\alpha$ marks an angle between $\mathbf{s}$ and $\mathbf{n}_0$. Black (a,b,e) and white [inset of (b)] thin arrows point to a bulk −1/2 defect loop (a) and to the −1/2 defect line branches (b,e), respectively. (j–n) Schematic diagrams showing $\mathbf{n}(\mathbf{r})$ with only edge-pinned surface defects (j,k), with both bulk and edge-pinned defect lines (l) and absorption of a bulk defect at the edge upon tilting the particle (m,n). Solid red lines and filled circles in (f–n) mark the bulk−1/2 disclinations. Solid orange and dark-blue lines and squares show the surface-bound ±1/4 defect lines.

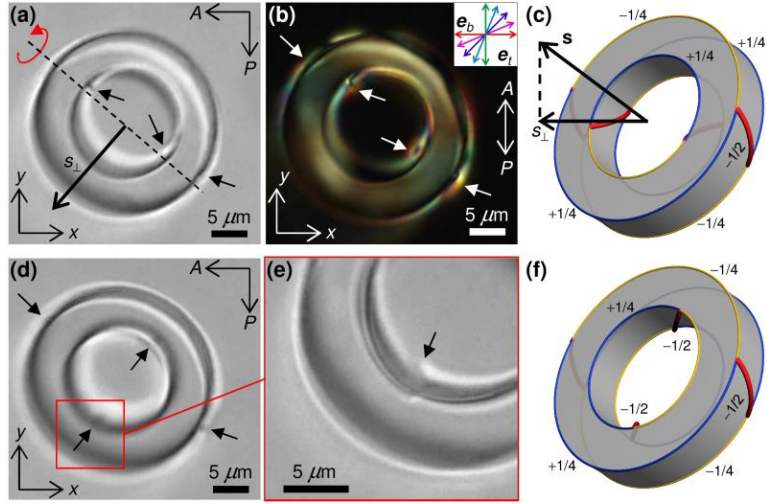

FIG. 2. Edge-pinned defect lines around a particle in a twisted nematic cell: (a,b,d,e) Optical micrographs obtained between crossed (a,d,e) and parallel (b) polarizers *P* and *A*. Orientation of a pair of bulk −1/2 defects (shown by black arrows) inside a particle's ring in (a) and (d) is different with respect to $s_\perp$. Inset in (b) shows right-handed twist of n(r) between bottom and top substrates with orthogonal easy axes $\mathbf{e}_b$ and $\mathbf{e}_t$, respectively. (e) Optical texture showing a zoomed-in view of a bulk −1/2 defect line seen in (d) while traversing from one edge to another. (c,f) Schematic diagrams of defects and their transformation corresponding to particles shown in (a,d), respectively. The handle-shaped bulk defect lines are shown using thick red lines. The quarter-strength edge-bound surface defect lines of opposite strengths are shown using thin blue and orange lines, with the strength of opposite signs marked next to them.

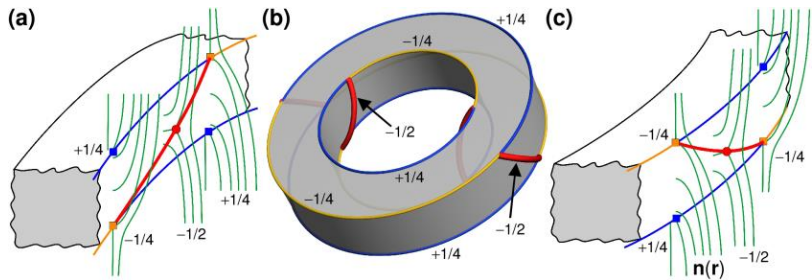

FIG. 3. Detailed schematic diagrams of **n(r)** and defects in the region of transformation of disclinations inside (a) and outside (c) of a toroidal particle with square-shaped cross section (b). The handle-shaped bulk defect lines are depicted as thick red lines. The quarter-strength edge-bound surface defect lines of opposite signs are shown using thin blue and orange lines, with the strength of opposite signs marked next to them. The green lines depict the director field.

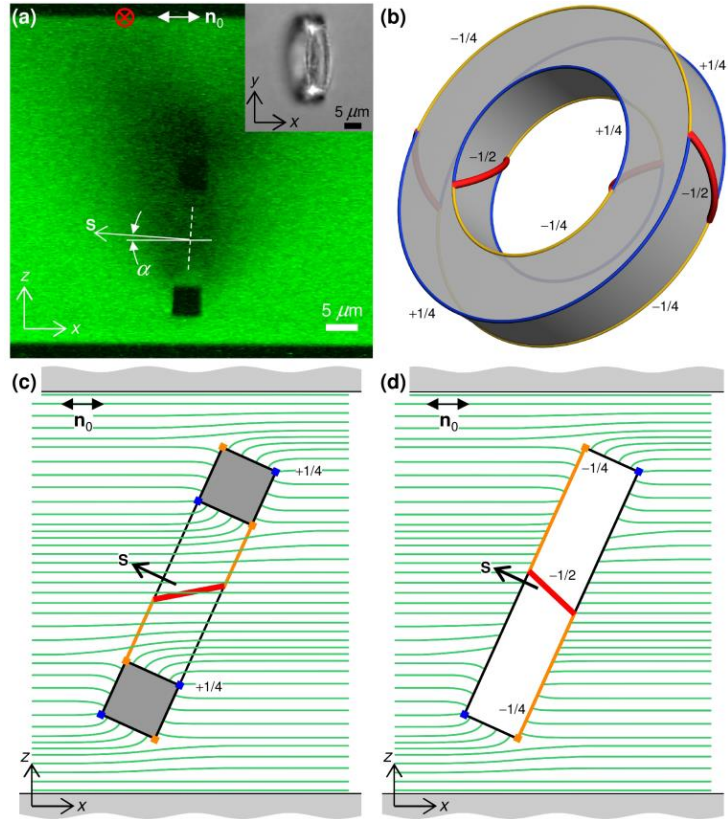

FIG. 4. Edge-pinned defect lines around a ring-shaped particle in a planar nematic cell: (a) a 3PEF-PM texture for a sample $xz$ cross section across the thickness of the cell showing that a torus is tilted with respect to the middle of the cell and is levitating in the bulk of the sample. The inset shows a bright-field micrograph of a tilted particle. (b) 3D schematic diagram of bulk and edge-pinned surface defects that undergo transformation. (c,d) Schematic diagram of $\mathbf{n}(\mathbf{r})$ and the induced defects inside (c) and outside (d) the particle. As in Fig. 2, the handle-shaped bulk defect lines are shown using thick red lines. The quarter-strength edge-bound surface defect lines of opposite strengths are shown using thin blue and orange lines, with the strength of opposite signs marked next to them. The green lines depict the director field.

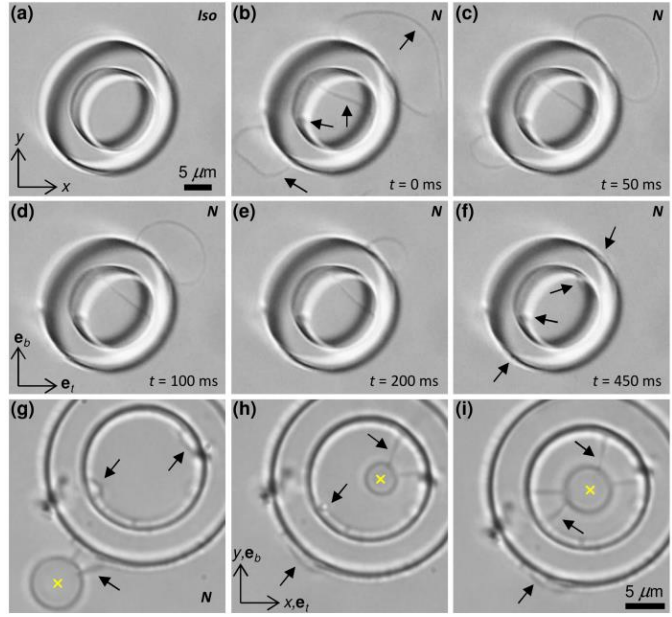

FIG. 5. (a–f) Nucleation, transformation, and absorption of bulk −1/2 defect lines at the edges of a tilted toroidal particle when locally quenching a nematic from an isotropic $I$ (a) to nematic $N$ (b–f) phase in a twist nematic cell upon switching laser tweezers off. The dark spot at the outer edge of the particle seen in (b–f) at around 9 o'clock corresponds to the debris attached to the particle below the top visible edge, which is not visible in the $I$ phase (a) because of a different particle tilt. (g–i) Optical trapping and stretching of the bulk −1/2 defects with laser tweezers in the exterior (g) and interior (h,i) of the particle's ring. Black arrows point to the bulk −1/2 defect lines. Yellow (bright) crosses in the middle of isotropic droplets of 5CB track the spatial position of a trapping laser beam when it is turned on.

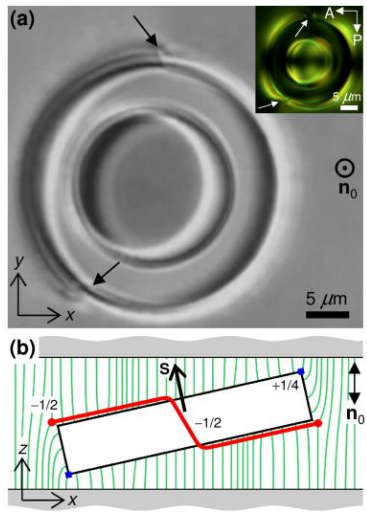

FIG. 6. Deformation of bulk disclinations around torus-shaped colloidal particles in a nematic LC. (a) Bright-field optical micrograph of a tilted particle with an outer bulk half-integer disclination that is not absorbed at the particle edges. Inset shows a corresponding polarizing microscopy texture obtained between crossed polarizers $P$ and $A$. Black and white (in the inset) thin arrows mark points where the deformed outer half-integer defect loop transitions between the top and bottom edges. (b) A schematic diagram showing the corresponding $\mathbf{n}(\mathbf{r})$.